\documentclass[twocolumn,showpacs]{revtex4}
\usepackage{epstopdf}
\usepackage{mathbbol}
\usepackage{amsmath}
\usepackage{graphics,graphicx,epsfig,ulem}
\begin{document}
\title{Electromagnetically induced transparency of an interacting cold Rydberg ensemble}
\date{\today}
\author{K. J. Weatherill, J. D. Pritchard, R. P. Abel, M. G. Bason, A. K. Mohapatra and C. S. Adams}

\affiliation{Department of Physics, Durham University, Rochester Building, South Road, Durham DH1 3LE, UK}

\begin{abstract}

We study electromagnetically induced transparency (EIT) of a weakly interacting cold Rydberg gas. We show that the onset of interactions is manifest as a depopulation of the Rydberg state and numerically model this effect by adding a density-dependent non-linear term to the optical Bloch equations. In the limit of a weak probe where the depopulation effect is negligible, we observe no evidence of interaction induced decoherence and obtain a narrow Rydberg dark resonance with a linewidth of $<$600~kHz, limited by the Rabi frequency of the coupling beam.
\end{abstract}
\maketitle

Ensembles of Rydberg atoms display fascinating many-body behavior due to their strong interactions \cite{gallagher}. These interactions lead to interesting cooperative effects such as superradiance \cite{gros82,rehl71,wang07} and dipole blockade \cite{sing04,tong04,cube05,heid07}, which may provide the basis for applications such as single--photon sources \cite{saff02} and quantum gates \cite{jaksch00,lukin01,saff05}.
For quantum information applications one is interested in the coherent evolution of the ensemble. Coherent excitation of Rydberg states has been achieved using adiabatic passage \cite{cube05b,deig06}. Rabi oscillations between ground and Rydberg states with dipole--dipole interactions have been observed \cite{john08,reet08}. Also, the coherence of a Rydberg ensemble has been measured directly using a spin echo technique \cite{rait08}. In most experiments on ultracold Rydberg gases, the Rydberg atoms are detected indirectly following field ionization and subsequent detection of electrons (or ions) using a micro channel plate (MCP). However, recently we demonstrated non-destructive optical detection of Rydberg states in room temperature Rb vapor \cite{moha07,baso08} using EIT \cite{harris,flei05}. The same technique was subsequently used to detect Rydberg states in a Sr atomic beam \cite{maug07}. Rydberg EIT has number of potential applications, for example, a Rydberg EIT medium displays a dc electro---optic effect many orders of magnitude larger than other systems \cite{moha08}, and Rydberg EIT enables direct measurement of the coherence of the Rydberg ensemble.

\begin{figure}[hbt]
\begin{center}
\epsfig{file=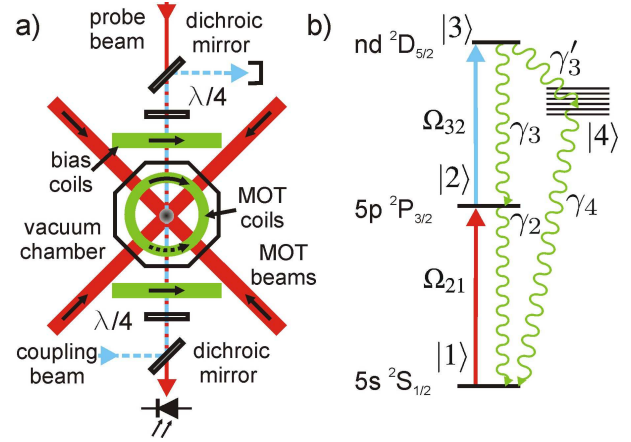,clip=,width=8.0cm}
\caption[]{(Color online) (a) Schematic of the experimental setup. A strong coupling beam (blue) stabilized to the 5p $^{2}$P$_{3/2} \rightarrow n$D (or $n$S) transition counter--propagates a weak probe beam (red) through a cold sample of $^{85}$Rb released from a magneto--optical trap (MOT). The probe scans over a range of $\pm$20~MHz about the $F=3 \rightarrow F'=4$ resonance in 480~$\mu$s. (b) Simplified level scheme, Rabi frequencies and decay rates used to model the Rydberg EIT. Levels $|4\rangle$ represent states other than state $|2 \rangle$ to which the Rydberg state $|3\rangle$ can decay.}
\end{center}
\end{figure}

In this work we demonstrate EIT involving Rydberg states in an ultra--cold atomic sample. We show that interactions between Rydberg atoms lead to a rapid depopulation of the Rydberg state, and that the EIT spectra are extremely sensitive to small changes in the interaction strength. For example, changing the principal quantum number of the Rydberg state from $n$ to $n+1$ produces a significant change in the EIT spectrum. By using a weak probe, where the probability of populating the Rydberg state is low, we can eliminate this depopulation effect and obtain narrow Rydberg dark resonances with a linewidth of $<~$600~kHz, limited by the Rabi frequency of the coupling beam.

The experimental setup and simplified level scheme are shown in figure 1 (a) and (b) respectively. A probe and coupling beam are combined using dichroic mirrors and counter--propagate through a cloud of laser--cooled $^{85}$Rb atoms. The polarization of the beams are chosen to maximize transition strengths ($\sigma^+$--$\sigma^+$). The probe beam is derived from a diode laser at 780.24 nm stabilized to the $\vert 1\rangle \rightarrow \vert 2\rangle$, 5s $^2$S$_{1/2}(F = 3)$ $\rightarrow$ 5p $^2$P$_{3/2}$ $(F' = 4)$ transition using polarization spectroscopy \cite{pear02}. The frequency is controlled using two successive acousto-optic modulators (AOMs). The beam is then coupled into an optical fiber, the output of which is collimated and passed through an aperture resulting in a beam of almost uniform intensity and radius of 0.75~mm. The coupling beam is derived from a commercial frequency doubled diode laser system (Toptica TA-SHG) at $\lambda$=483--487~nm resonant with the $\vert2\rangle \rightarrow \vert3\rangle$, 5p $^2$P$_{3/2}$ $(F' = 4)$ $\rightarrow$ $n$d $^2$D$_{5/2}$ (or $n$s $^2$S$_{1/2}$) $(F'')$ transition. The frequency of the coupling laser is stabilized using a Rydberg EIT signal produced in a magnetically shielded rubidium vapor cell at room temperature \cite{moha07}.

We load approximately $10^8$ atoms in a magneto--optical trap (MOT) in 5~s. A subsequent 10~ms optical molasses phase cools the atoms to a temperature of approximately 20~$\mu$K and density
$10^{10}~{\rm cm}^{-3}$, following which the cooling light is extinguished. An optical pumping pulse of variable duration prepares the atoms in the $F=3$, $m_F=3$ state. After 3~ms free expansion the pump and probe beams are turned on and the frequency of the probe is swept across the D2 $F=3 \rightarrow F'=4$ transition. The sweep covers $\pm$ 20 MHz about the resonance in 480~$\mu$s. Absorption measurements show that there is negligible density decrease over the duration of the scan and negligible optical pumping into the lower ($F=2$) hyperfine ground state for the probe powers of interest. The probe beam is incident on a fast photodiode which records the transmission through the sample. This technique of measuring the EIT spectra from a cold atomic sample within a single experimental cycle offers a fast repetition rate.

In figure~2~ we show a series of EIT spectra for a fixed probe power of 3.6~$\mu$W corresponding to 0.13~$I_{\rm sat}$, where $I_{\rm sat}$ is the saturation intensity on the $\vert 1\rangle \rightarrow \vert 2\rangle$ transition.  The absorption data has been corrected for the measured change in probe intensity through the optical fiber as a function of AOM frequency. The effect of varying atom density, $\rho$, and principal quantum number, $n$, are shown in figures 2 (a) and (b), respectively. The atom density, $\rho$, is changed by heating the atom cloud (of initial density $\rho_0$) with the probe beam prior to the EIT scan. In figure 2 (b) the power of the coupling laser is varied to make the coupling Rabi frequency the same for each scan. For low $\rho$ or low $n$ a characteristic EIT spectrum is observed with a slight asymmetry. As $\rho$ or $n$ are increased this asymmetry becomes more pronounced until the transparency peak is lost completely. The lineshape for $n>22$ does not change dramatically with $n$ as illustrated by the similarity to the $n=26$ and $n=22$ data sets, the front curves in figure 2 (a) and (b), respectively. The data presented in figure 2 is suggestive of a density--dependent depopulation of the Rydberg state.

\begin{figure}[hbt]
\begin{center}
\epsfig{file=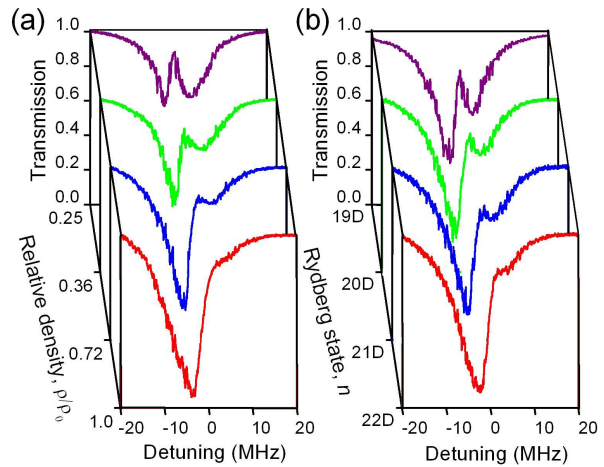,clip=,width=8.0cm}
\caption[]{(Color online) (a) Transmission of a 3.6~$\mu$W probe beam as a function of probe laser detuning for the $^{85}$Rb $|1\rangle \rightarrow |2\rangle \rightarrow |3\rangle=26{\rm d}$ system for varying atomic density.  (b) EIT spectra for $n$=19--22. The Rabi frequency of the $|2\rangle \rightarrow |3\rangle$ transition is kept the same for each state. In both plots the onset of density--dependent behavior is evident in the emergence of an EIT lineshape at lower values of $n$ and $\rho$} \end{center}
\end{figure}

By varying the probe power we can change the fractional probability that an atom is excited to the Rydberg state. In figure 3 we show the transmission spectrum for the $\vert3\rangle$=19d state for two probe powers of 450~nW (0.016~$I_{\rm sat}$) and 3.6~$\mu$W (0.13~$I_{\rm sat}$). In this case we present double scans where the probe laser detuning is first increased and then decreased through the resonance in a total time of 960~$\mu$s. We observe that at the higher probe power there is significantly less absorption on the second scan than the first indicating that atoms excited to the Rydberg state are being lost from the EIT system. For these parameters only a small fraction of the population, a few percent, is transferred into and out of the Rydberg state for a time of order 10~$\mu$s.  However, to model the data (see next paragraph) we find that $>$70$\%$ of the atoms are lost from the 3-level EIT system. This substantial loss cannot be explained by spontaneous decay from Rydberg state (since the spontaneous lifetime is also of order $\sim$10~$\mu$s \cite{theo84} and decreases with larger $n$) or by optical pumping to the lower hyperfine state. We therefore deduce that there is an interaction induced loss from the system happening on a much faster time scale than the Rydberg state lifetime.

To describe this interaction--induced depopulation we add non-linear density dependent decay and decoherence terms to the standard three-level
optical Bloch equations \cite{flei05}:
\begin{equation}
i\hbar\dot{\sigma} = [H,\sigma] - \gamma\sigma
\end{equation}
where $\sigma$ is the density matrix, $H$ the atom-light interaction Hamiltonian and $\gamma\sigma$ the decoherence matrix. The depopulation of the Rybderg state is modeled by introducing a fourth level $\vert4\rangle$ to represent the reservoir of states other than $\vert2\rangle$ to which an atom in Rydberg state $\vert3\rangle$ could decay, as shown in figure 1 (b). The decay of the Rydberg state $\vert3\rangle$ is given by $\gamma_{3}\sigma_{33} +\gamma_{3}^{'}\sigma_{33}^{2}$ where the first term is due to spontaneous emission and the second term is a density-dependent interaction induced loss to the reservoir states $\vert4\rangle$. The spontaneous decay rates of states $\vert 2\rangle$ and $\vert3\rangle$ are given by $\gamma_{2}$ and $\gamma_{3}$, respectively. We assume that 50$\%$ of the spontaneous decay from $\vert3\rangle$ populates state $\vert 2\rangle$ and 50$\%$ is lost to the reservoir $\vert 4\rangle$, however this weighting makes little difference to the lineshape as it is the non-linear term, $\gamma_{3}^{'}$, that dominates. Atoms decay from $|4\rangle $ to $|1\rangle$ at a rate $\gamma_{4}$. The coherences are assumed to decay at half the rate of the populations. The decoherence rate between $|1\rangle$ and $|3\rangle$, $\gamma_{31}$, is included as a fitting parameter which depends on the combined laser linewidth and dephasing due to Rydberg atom interactions. The optical Bloch equations are then solved using a time varying probe detuning to match the experimental sequence. As the probe is swept through the two--photon resonance the Rydberg state is first populated and then depopulated by chirped adiabatic passage, in contrast to conventional adiabatic passage schemes where the transfer is unidirectional \cite{cube05b,deig06}.
The predictions of the model are compared to the experimental spectra in figure 3. The model accurately reproduces the observed spectra for different values of the probe laser intensity (all other parameters remain the same). When the probe intensity is increased the population of the Rydberg state (solid lines in the lower panel of figure 3) also increases and there is a faster decay into the reservoir states (dotted line in the lower panel of figure 3). For a higher probe power of 3.6~$\mu$W the model predicts that about 3$\%$ of the population is transferred into the Rydberg state and after the double scan sequence almost 80$\%$ of the population is in other states, $|4\rangle$.
\begin{figure}[hbt]
\begin{center}
\epsfig{file=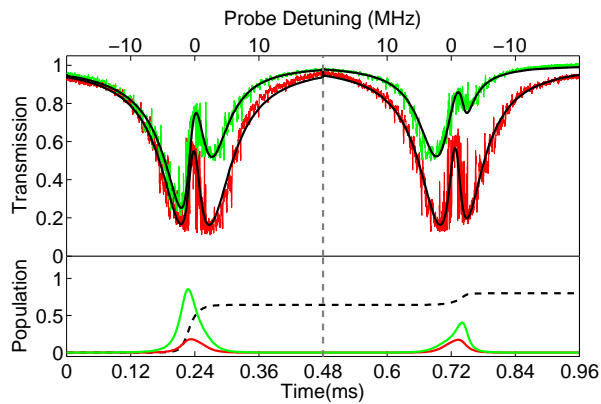,clip=,width=8.0cm}
\caption[]{(Color online) Top: Double scan EIT for the $\vert1\rangle \rightarrow \vert2\rangle \rightarrow \vert3\rangle =19$d system for probe powers of 450~nW (red) and 3.6~$\mu$W (green). The model fits are shown as the solid black lines. Bottom: The populations of the Rydberg states $\vert3\rangle$ $\times 25$ (solid) and reservoir states $|4\rangle$ (dashed) as a function of time during the scan}
\end{center}
\end{figure}

There are a number of surprising features of this density--dependent depopulation effect. First, we observe similar decay rates for nearby s--states. Second, the observed decay is relatively fast for such low Rydberg atom densities of order 10$^8$~cm$^{-3}$. For example, the time scale associated with near--resonant dipole--dipole energy transfer processes \cite{tann08} at this density would be 10 to 20 times slower. Also, our simple non--linear optical Bloch equation model begins to fail for larger $n$. For example, for the high density 26d trace shown in figure 2 (a) no evidence of EIT is observable whereas the model predicts a pronounced shoulder after the laser has passed through the two--photon resonance. One possible explanation is that the decay out of the EIT system is driven by superradiant decay of the Rydberg state to nearby f and p states. The potential importance of superradiance in frozen Rydberg gases particularly involving lower $n$ states has been highlighted in recent work \cite{wang07,day08}. In general superradiance is more significant for low $n$ as the decay rate to nearby Rydberg levels scales as $1/n^5$, however there is also a geometrical effect which depends on the wavelength of the superradiant transition, $\lambda_{\rm s}$, in comparison to the size of the cloud, $D$ \cite{day08}. For the range of $n$ studied in this work ($n=19$ to 26) $\lambda_{\rm s}$ is less than $D$ for decay to the closest f states i.e. the atoms are spatially spread over a range greater than the transition wavelength such that cooperative effects are small. However, $\lambda_{\rm s}$ is of order or greater than $D$ for decay to the closest p states i.e. all atoms are within one wavelength and the emission rate is greatly enhanced. In this critical region the geometric cooperativity parameter \cite{day08} increases rapidly with increasing $n$ which could explain the dramatic evolution of EIT spectra shown in figure 2 (b). The dependence of superradiance on $\lambda_{\rm s}$ and $D$ is supported by the similarity of figure 2 (a) and (b) where for both data sets the value of $\lambda_{\rm s}D$ varies by a factor of 1.6 between the front and back trace \cite{footnote}. Finally, we note that one can obtain a better fit to
the high density and higher $n$ data by coupling the optical Bloch equations to an equation describing the superradiance field energy density. This model reproduces the lineshapes observed in the experiment, in particular the complete loss of the EIT shoulder, however as the time scale of the decay of the superradiance field is not known,
this model does not provide useful quantitative information.

Lastly, we focus on the weak probe limit where the population of the Rydberg state becomes negligible. This limit highlights the power of EIT as a non-destructive probe of Rydberg gases.
Figure 4 shows the transmission spectrum for the $\vert 3\rangle$=26d state using a low probe power of 200~nW (0.007~$I_{\rm sat}$). In this case no loss or additional decoherence of the Rydberg state is observed between the forward and back scan. We obtain a typical EIT linewidth of 600~kHz. The EIT linewidth is made up of two terms; a term proportional to the coupling Rabi frequency squared, $\Omega_{32}^2$ (i.e. the Autler--Townes splitting), and a term corresponding to the coherence between the ground state and the Rydberg state, $\gamma_{31}$. This coherence term is the sum of the combined laser linewidth and interaction dependent dephasing mechanisms. From the fit we obtain $\Omega_{32}/2\pi=1.8\pm 0.3$~MHz and $\gamma_{31}=100\pm 20$~kHz, i.e. for our parameter the linewidth is dominated by the Autler-Townes term. We observe no change in $\gamma_{31}$ over the range $n=19$ to $n=26$ showing that the combined laser linewidth dominates over any interaction induced dephasing of the Rydberg state. Note that such long coherence times cannot be observed in non--Rydberg cascade EIT due to the short lifetime of the upper state. The frequency resolution obtained using this method is comparable to the best obtained on single atoms \cite{john08} and considerably better than the resolution obtained for other ensembles experiments \cite{grab06,reet08,vogt06}, where interaction effects tend to broaden the linewidth to greater than one MHz. This narrow resonance demonstrates a significant advantage of EIT for precision spectroscopy or electrometry, i.e. that the observed transparency is sensitive to the Rydberg energy level without transferring significant population into it. From the model we find that the probability for an atom to be in the Rydberg state to be $\sim$0.6$\%$ giving $\rho< 10^7$~cm$^{-3}$ which corresponds to a total Rydberg population of less than 10$^4$ atoms.

\begin{figure}[hbt]
\begin{center}
\epsfig{file=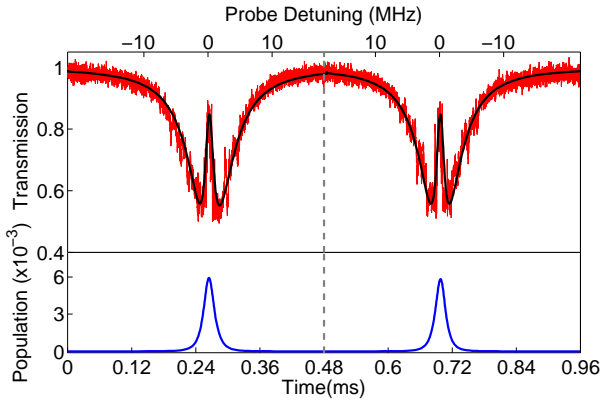,clip=,width=8.0cm}
\caption[]{(Color online) Top: Double scan EIT for the $|1\rangle \rightarrow \vert2\rangle \rightarrow |3\rangle=26{\rm d}$ system. The solid line (black) is the theoretical line of best fit giving an EIT linewidth of 0.58 $\pm$ 0.04~MHz. Bottom: Population of the Rydberg state as a function of time for the duration of the scan.}
\end{center}
\end{figure}

In conclusion, we have shown that EIT is a useful technique to study the properties of interacting Rydberg ensembles. We find that interaction effects are manifest as a rapid depopulation of the Rydberg state. This density dependent loss can be explained using a simple non--linear optical Bloch equation model. The character of the loss is similar for changes in either the atomic density or the principal quantum $n$ of the Rydberg state, suggesting a superradiant decay mechanism. In the limit of weak probe power, where the depopulation effect is negligible we observe no evidence of interaction induced decoherence and the linewidth is limited by the Autler--Townes splitting. In this limit EIT provides a sensitive non--destructive probe of a cold Rydberg ensemble allowing high--resolution spectroscopy of Rydberg states.

We are grateful to PF Griffin for experimental assistance, IG Hughes, MPA Jones, T Pfau, and TG Walker for stimulating discussions and to SL Cornish for the loan of equipment. We thank the EPSRC for financial support.


\begin{thebibliography}{99}
\bibitem{footnote} In figure~2~(a) the density changes by a factor of 4 and hence the sample size $D$ by a factor of $\sqrt[3]{4}=1.58$ (with $\lambda_{\rm s}$ constant). In figure 2 (b) the cloud size and density remain constant but $n$ changes from 19 to 22 and hence $\lambda_{\rm s}$ changes by a factor of $22^3/19^3=1.55$.
\end{thebibliography}
\end{document}